\documentclass[a4paper,11pt]{article}
\usepackage{jheppub} 
\usepackage{lineno}
\usepackage{bm}
\usepackage{booktabs}
\usepackage{subcaption}
\usepackage{placeins}

\title{\boldmath Impact of Heavy Modes on Primordial Black Hole Formation}

\author[a]{Guo-He Li}
\author[a]{Mian Zhu}
\author[b]{Chunshan Lin}

\affiliation[a]{College of Physics, Sichuan University, Chengdu 610065, China}
\affiliation[b]{Faculty of Physics, Astronomy and Applied Computer Science, Jagiellonian University, 30-348 Krakow, Poland}

\emailAdd{zhumian@scu.edu.cn}

\abstract{%
	Primordial black holes (PBHs) form from rare fluctuations in the high-density tail of the primordial distribution, making their abundance highly sensitive to primordial non-Gaussianity. Heavy fields during inflation can source such non-Gaussianity, producing characteristic signatures widely explored in cosmological collider physics, yet its impact on PBH formation remains largely unexamined. We investigate this question within quasi-single-field inflation (QSFI), the minimal realization that couples a heavy field to the inflaton. Unlike most PBH studies that consider only the bispectrum, the QSFI can generate a sizable trispectrum alongside the bispectrum, allowing us to investigate their combined effects on PBH formation. In our leading-order calculation, heavy-field non-Gaussianities can amplify the high-density tail and substantially enhance PBH production. Within the parameter scan, the largest enhancement occurs at the smallest heavy-field mass and the largest sound-speed ratio considered.  The trispectrum contribution can be comparable to that of the bispectrum and may even dominate the total non-Gaussian correction. Our results establish PBH abundance as a novel observational channel for cosmological collider physics, demonstrating how heavy fields during inflation can significantly shape PBH formation.
}

\begin{document}
	\maketitle
	\flushbottom

	\section{Introduction}
	\label{sec:intro}
	
	Primordial black holes (PBHs), formed via the gravitational collapse of overdense regions sourced by large primordial density fluctuations in the primordial universe \cite{Zeldovich:1967lct,Hawking:1971ei,Carr:1974nx}, can span a wide mass range from micrograms to millions of solar masses. Consequently, PBHs have attracted considerable attention \cite{Zeng:2024snl,Ning:2022lww,Yu:2026vey,Mohammadi:2025avz,Dave:2026gsd,Zhong:2025xwm,He:2024luf,Papanikolaou:2024fzf,Cai:2023ptf,Sun:2025yiq,Zhai:2023azx,Fu:2020lob,Liu:2021jnw,Liu:2020cds,LHAASO:2025kyn,Gu:2023mmd,Tao:2026ltc} and are relevant to a variety of astrophysical and cosmological phenomena such as the origin of dark matter \cite{Carr:2016drx,Carr:2020xqk,Xie:2024eug} and the seeds of supermassive black holes (BHs) \cite{Kawasaki:2012kn,Liu:2022bvr, Huang:2023chx}.
	
	Given their importance, the origin of PBHs has been extensively investigated in the literature \cite{Kawasaki:2012wr,Chen:2016kjx,Quintin:2016qro,Wang:2016ana,Pi:2017gih,Nakama:2018utx,Ding:2019tjk,Domenech:2020ssp,Kawana:2021tde,Lin:2021vwc,Tan:2022lbm,Domenech:2021wkk,Liu:2021svg,He:2022amv,DeLuca:2022bjs,Lewicki:2023ioy,Zhang:2023tfv,Wang:2024nmd,Han:2026ycn,Gu:2022pbo,Cai:2022erk,Chen:2019zza,Liu:2019lul,Gao:2018pvq}. The most studied mechanism of PBH formation is the collapse of overdense regions due to self-gravity effect. These overdense regions originate from the classicalization of quantum fluctuations in the primordial universe, and their energy density must exceed a certain threshold for collapse into BHs to occur. The curvature fluctuations on the PBH formation scale must therefore be orders of magnitude larger than those on Cosmic Microwave Background (CMB) scales \cite{Sasaki:2018dmp}. However, canonical inflationary scenarios predict a nearly scale-invariant power spectrum of curvature fluctuations, which is insufficient for PBH formation. Viable PBH production thus requires the curvature power spectrum to be enhanced up to $\mathcal{P}_{\zeta} \sim 10^{-2}$ on small scales, roughly seven orders of magnitude above its CMB value, and abundant efforts have been devoted to realizing such an enhancement within inflation \cite{Motohashi:2017kbs,Hertzberg:2017dkh,Germani:2018jgr,Garcia-Bellido:2017mdw,Germani:2017bcs,Byrnes:2018txb,Di:2017ndc,Ragavendra:2020sop,Dalianis:2018frf,Cai:2018tuh,Cai:2019bmk,Zhou:2020kkf,Cai:2020ovp,Cai:2023uhc,Fu:2019ttf,Chen:2025qyv}. 
	
	Enhancing the power spectrum is, however, only part of the story. The statistical properties of curvature fluctuations are fully determined by the power spectrum only if their statistics are purely Gaussian, which is generally not the case in PBH formation scenarios. In fact, many such scenarios predict large non-Gaussianity on small scales \cite{Sasaki:2006kq,Fonseca:2011aa,Pi:2022ysn, Gow:2023zzp,Chen:2022usd}, which modifies the tail of the probability distribution function (PDF) of curvature fluctuations \cite{Niemeyer:1997mt,Shibata:1999zs,Musco:2004ak}, and can change the PBH abundance exponentially, see e.g., \cite{Pi:2024lsu} for a review. Non-Gaussianity is therefore an indispensable ingredient in any quantitative prediction of the PBH abundance. Moreover, non-Gaussianity may enhance the PBH abundance at a fixed amplitude of the curvature power spectrum compared to the Gaussian case  \cite{Seery:2006wk,Young:2013oia,Atal:2019cdz,Yoo:2019pma, Riccardi:2021rlf,Kitajima:2021fpq,Pi:2021dft,Young:2022phe,Kehagias:2019eil}. Equivalently, to generate a given PBH abundance, the power-spectrum amplitude need not reach $\mathcal{P}_{\zeta} \simeq 10^{-2}$ in the presence of non-trivial non-Gaussianity. A lower required amplitude may help evade the back-reaction problem associated with an oversized curvature power spectrum \cite{Kristiano:2022maq}, and circumvent the constraints on massive PBH formation from the CMB $\mu$-distortion limits on  $\mathcal{P}_{\zeta}$ at scales around $10^4 {\rm Mpc}^{-1}$ scales \cite{Fixsen:1996nj,Fixsen:1998kq}. 
	
	A common concern in such investigations is that the predicted non-Gaussianity depends significantly on the details of model building. In most PBH formation scenarios, the action is engineered to enhance the power spectrum on specific scales, so that the resulting non-Gaussian signatures are heavily model dependent, offering limited organizing principle or predictive power.
	
	There exists, however, a class of primordial non-Gaussianity whose ``model dependence'' is not a drawback but the very point. Massive fields are naturally present during inflation and can leave characteristic non-analytic signatures in the correlation functions of curvature perturbations, which systematically encode the masses and spins of the intermediate particles \cite{Chen:2009we,Chen:2009zp,Chen:2012ge,Arkani-Hamed:2015bza,Chen:2015lza,Lee:2016vti,Chen:2018xck}. This is the program of cosmological collider physics, which aims to develop systematic methods for extracting such signatures \cite{Arkani-Hamed:2018kmz,Meerburg:2016zdz,Lu:2021wxu,Jazayeri:2023xcj} and to turn the inflationary universe into a probe of new physics at the Hubble scale. The systematic nature of these heavy-field signatures raises a natural question: can the non-Gaussian correlations from heavy particles not only serve as cosmological collider signals, but also reshape the rare high-density tail of the probability distribution and thereby facilitate PBH formation? If so, the PBH abundance would become a novel observational channel for cosmological collider physics.
	
	In this paper, we investigate this question within quasi-single-field inflation (QSFI) \cite{Chen:2009we,Chen:2009zp,Chen:2012ge,Baumann:2011nk,Noumi:2012vr,Emami:2013lma,An:2017hlx,Iyer:2017qzw}, the simplest realization of the cosmological collider setup. In QSFI, a massive scalar field ($m \sim H$) is present alongside the inflaton, and the mixing between the two fields results in a turning background trajectory. The interactions between the heavy field and the inflaton generate a characteristic set of primordial correlators: a squeezed bispectrum, together with contact and scalar-exchange trispectra. Interestingly, the sizable primordial trispectrum is a feature in QSFI, absent in most inflationary scenarios. Therefore, the investigation of QSFI PBH formation goes beyond the bispectrum-only treatments adopted in most PBH studies involving non-Gaussian fluctuations \footnote{There are, of course, literatures on the impact of trispectrum \cite{Tada:2015noa,Matsubara:2019qzv,Suyama:2019cst,Matsubara:2022nbr} in PBH formation scenario.}.
	
	We find that, within the perturbatively controlled regime, heavy-field non-Gaussianity can amplify the high-density tail and substantially enhance PBH production. For instance, for a sample abundance $\beta_{\ast} = 10^{-15}$, the required peak amplitude can be reduced to about $59\%$ of the Gaussian value. Equivalently, at a fixed power-spectrum peak $\mathcal{P}_{\zeta} (k_{\ast}) = 10^{-2}$, the PBH abundance can be enhanced by more than twenty-nine orders of magnitude relative to the Gaussian prediction. Notably, the sizable heavy-field trispectrum contributes significantly to this enhancement. These results establish the PBH abundance as a novel observational channel for cosmological collider physics, and the possibilities of intermediate heavy fields to assist PBH formation.
	
	The paper is organized as follows. In Sec. \ref{sec:pbh_statistics}, we review the PBH abundance in Gaussian and non-Gaussian cases. We then introduce the QSFI framework and review its primordial bispectrum and trispectrum in Sec.~\ref{sec:qsfi}. The impact of heavy-field non-Gaussianity on the PBH abundance is investigated in Sec.~\ref{sec:qsfi_pbh_abundance}. We present our conclusions in Sec. \ref{sec:conclusions}.
	
	\section{The PBH formation mechanism}
	\label{sec:pbh_statistics}
	
	A local overdense region, by definition, has an energy density larger than the background value. We use density contrast, $\delta \equiv \rho/\bar{\rho} - 1$, where $\rho$ and $\bar{\rho}$ are the energy density of the overdense region and the cosmic background, to measure such difference. In the radiation dominated epoch, this quantity is related to the curvature fluctuations $\zeta$ in the Fourier space as \cite{Liddle:2000cg,Green:2004wb,Young:2014ana}
	\begin{equation}
		\delta_k(t) = \mathcal{M}(k) \zeta_k ~,~ \mathcal{M}(k) \equiv \frac{4}{9} \left(\frac{k}{aH}\right)^{2} ~,
	\end{equation}
	where we used the fact that the equation-of-state parameter $w = 1/3$ for radiations. The gravitational collapse takes place when the primordial fluctuations $\zeta_k$ re-enters the Hubble horizon (i.e., $k = aH$) and the overdense region becomes causally connected. The characteristic length scale for PBH formation is thus $R \simeq (aH)^{-1}|_{\rm form}$. 
	
	The gravitational collapse process shall be described in configuration space instead of momentum space, so work with the smoothed density contrast
	\begin{equation}
		\delta_R(\bm{x}) = \int \frac{d^3 k}{(2\pi)^3} e^{i\bm{k} \cdot \bm{x}} W(kR) \delta_k = \int \frac{d^3 k}{(2\pi)^3} e^{i\bm{k}\cdot\bm{x}} W(kR) \mathcal{M}(k) \zeta(\bm{k}) ~,
		\label{eq:DeltaR_zeta}
	\end{equation}
	where we adopt the Gaussian window $W(kR)=\exp[-(kR)^2/2]$ to remove modes far shorter than the smoothing scale \cite{Ando:2018qdb}. The overdense region has to be dense enough to collapse into a BHs. The peak theory suggests that $\delta_R$ must exceed a threshold value $\delta_c$ for PBH formation, i.e., $\delta_R \geq \delta_c$ \cite{Musco:2020jjb}. Nevertheless, it is argued that peak theory may not be the correct description for PBH formation in recent days \cite{Harada:2015yda,Escriva:2019phb,Yang:2024snb}. In this paper, we are interested in the impact of heavy fields on PBH formation, so we will put aside such debate and simply adopt $\delta_c = 0.5$ for illustrative purpose, since the peak theory suggests $0.4 < \delta_c < 0.7$. 
	
	Now, we shall evaluate how many overdense regions in the universe exceed such threshold. To do so, we need to know the one-point probability density function of $\delta_R$, $P(\delta_R)$ \cite{Sasaki:2018dmp}, which specifies the probability density associated with each possible value of the smoothed density contrast $\delta_R$. The PBH abundance is then related to the probability weight above the threshold $\delta_c$. In particular, the standard Press-Schechter formalism \cite{Press:1973iz} tells
	\begin{equation}
		\beta \equiv \left. \frac{\rho_{\rm PBH}}{\rho_{\rm tot}}\right|_{\rm form}
		= \gamma \int_{\delta_c}^{\infty} P(\delta_R) d\delta_R ~,
		\label{eq:beta_general}
	\end{equation}
	where the mass fraction function $\beta(M)$ describes the fraction of PBHs
	compared to the total energy of the universe and characterize the abundance of PBHs, and the collapse efficiency is taken to be $\gamma=0.2$ \cite{Carr:1974nx}. 
	
	Here's how the statistics of primordial fluctuations enters into the physics. When the curvature fluctuations are Gaussian, the one-point probability distribution is simply
	\begin{equation}
		P_{\rm G}(\delta_R)	= \frac{1}{\sqrt{2\pi} \sigma}\exp \left(-\frac{\delta_R^2} {2\sigma^2} \right) ~,
		\label{eq:PG_delta}
	\end{equation}
	with variance
	\begin{equation}
		\sigma^2(R) \equiv \left \langle \delta_R^2 \right\rangle =	\int \frac{dk}{k}	W^2(kR)	\frac{16}{81}(kR)^4	\mathcal{P}_\zeta(k),
		\label{eq:sigma_delta_rad}
	\end{equation}
	where $\mathcal{P}_\zeta(k)$ is the dimensionless curvature power spectrum. We immediately see that in Gaussian case, the PBH abundance is totally determined by the curvature power spectrum.
	
	When non-Gaussian statistics are taken into account, \eqref{eq:PG_delta} no longer holds. Specifically, in the Gaussian case, the PBH abundance in the high-threshold regime $\nu_{\rm th} \equiv \delta_c/\sigma \gg 1$ reduces to 
	\begin{equation}
		\beta_{\rm G} \simeq \frac{\gamma}{\sqrt{2\pi} \nu_{\rm th}} \exp \left(-\frac{\nu_{\rm th}^2}{2}\right) ~.
		\label{eq:beta_Gaussian}
	\end{equation}
	where $\beta_{\rm G}$ denotes the PBH abundance evaluated under Gaussian statistics. As a result, even a small modification of $\nu_{\rm th}$ from non-Gaussian statistics may lead to a significant change of PBH abundance. We are thus motivated to investigate the non-Gaussian effects on PBH formation.
	
	To see this more clearly, we follow \cite{Matsubara:2022nbr} by expanding the one-point PDF in an Edgeworth series and isolating the leading asymptotic terms in the high-peak limit $\nu_{\mathrm{th}} \equiv \delta_c/\sigma \gg 1$, which yields~\cite{Matsubara:2022nbr}
	\begin{equation}
		\beta_{\rm NG} \simeq \beta_{\rm G}	\exp\left[\nu_{\rm th}^2\sum_{n=3}^{\infty}\frac{\delta_c^{n-2}}{n!}S_n\right] =	\frac{\gamma}{\sqrt{2\pi}\,\nu_{\rm th}} \exp\left[ -\frac{\nu_{\rm th}^2}{2} \left( 1-2\sum_{n=3}^{\infty}\frac{\delta_c^{n-2}}{n!}S_n \right) \right] ~,
		\label{eq:beta_NG_all_cumulants}
	\end{equation}
	where $\beta_{\rm NG}$ denotes the PBH abundance with non-Gaussian corrections, and the reduced cumulants are defined as
	\begin{equation}
		S_n \equiv \frac{\langle \delta_R^n\rangle_c}{\sigma^{2n-2}} ~,~ n\geq 3 ~,
		\label{eq:Sn_def}
	\end{equation}
	where $\langle \delta_R^n\rangle_c \propto \sigma^{2n-2}$. It is clear that the cumulants carry the information on non-Gaussian effects of PBH formation. In addition, we see $S_n$ is independent of $\sigma$ in the hierarchical sense, i.e. at fixed template-normalized non-Gaussian amplitudes. Since $\delta_R$ is related to $\zeta_k$ via \eqref{eq:DeltaR_zeta}, $S_n$ can be extracted from the n-point correlation functions of curvature fluctuations. For example, the bispectrum $B_\zeta$ and trispectrum $T_\zeta$ are defined by
	\begin{align}
		\langle \zeta(\bm{k}_1)\zeta(\bm{k}_2)\zeta(\bm{k}_3)\rangle_c &=	(2\pi)^3\delta_D^{(3)}(\bm{k}_1+\bm{k}_2+\bm{k}_3)	B_\zeta(\bm{k}_1,\bm{k}_2,\bm{k}_3) ~, \label{eq:Bzeta_def} \\
		\langle \zeta(\bm{k}_1)\zeta(\bm{k}_2)\zeta(\bm{k}_3) \zeta(\bm{k}_4)\rangle_c &=	(2\pi)^3\delta_D^{(3)}(\bm{k}_1+\bm{k}_2+\bm{k}_3+\bm{k}_4) T_\zeta(\bm{k}_1,\bm{k}_2, \bm{k}_3,\bm{k}_4) ~,
		\label{eq:Tzeta_def}
	\end{align}
	we immediately see that the bispectrum and trispectrum are associated to $S_3$ and $S_4$ via
	\begin{align}
		& S_3 \equiv \frac{\langle \delta_R^3\rangle_c}{\sigma^4} = \frac{1}{\sigma^4} \int_{\bm{k}_{123}=0} \prod_{i=1}^{3}\left[\mathcal{M}(k_i)W(k_iR)\right]	B_\zeta(\bm{k}_1,\bm{k}_2,\bm{k}_3) ~,
		\label{eq:S3_general} \\
		& S_4 \equiv \frac{\langle \delta_R^4\rangle_c}{\sigma^6} = \frac{1}{\sigma^6} \int_{\bm{k}_{1234}=0} \prod_{i=1}^{4}\left[\mathcal{M}(k_i)W(k_iR)\right]	T_\zeta(\bm{k}_1,\bm{k}_2,\bm{k}_3,\bm{k}_4) ~, \\
		& \int_{\bm{k}_{1\dots n}=0} \equiv \prod_{i=1}^n \frac{d^3k_i}{(2\pi)^3} (2\pi)^3\delta_D^{(3)}(\sum_i\bm{k}_i) ~.
		\label{eq:S4_general}
	\end{align}
	Cosmologists conventionally work with the cases where the bispectrum and trispectrum provide the dominant non-Gaussian contribution. Eq.~\eqref{eq:beta_NG_all_cumulants} then simplifies to
	\begin{equation}
		\beta_{\rm NG} \simeq \beta_{\rm G} \exp\left[ \nu_{\rm th}^2\delta_c	\left(\frac{S_3}{6}+\frac{\delta_c S_4}{24} \right) \right] ~.
		\label{eq:beta_NG_S3S4}
	\end{equation}
	Therefore, a positive value of the cumulant correction can increase the PBH abundance exponentially. The quantities $S_3$ and $S_4$, named as the skewness/kurtosis in Ref. \cite{Matsubara:2022nbr}, control such amplifications, thus being our primary goals to compute.
	
	In the end, we comment on the validity of perturbative expansions. The expansion \eqref{eq:beta_NG_all_cumulants} holds only if the cumulant correction remains perturbative. A sufficient condition is that the correction does not reverse the sign of the leading exponential suppression,
	\begin{equation}
		2\sum_{n=3}^{\infty} \frac{\delta_c^{n-2}}{n!} S_n < 1 ~,
		\label{eq:consistency_condition}
	\end{equation}
	With our threshold choice $\delta_c = 0.5$, this requirement simplifies to
	\begin{equation}
		Q_{\rm NG} \equiv S_3 + \frac{S_4}{8} \lesssim 6 ~,
	\end{equation}
	In the following, we strictly impose the perturbative control bound $Q_{\rm NG} \lesssim 6$ for all numerical calculations.

	\section{Quasi-Single Field Inflation}
	\label{sec:qsfi}
	
	\subsection{The model}
	\label{subsec:qsfi_sound_speeds}
	QSFI takes into account the impact of massive fields with a mass of order the Hubble scale $H$. In the minimal setup, the massive modes modify the background evolution to a slowly turning trajectory in field space with a radius $\rho$ and a constant turning rate. The matter action is \cite{Chen:2009we,Chen:2009zp} 
	\begin{equation}
		S_m = \int d^4x\sqrt{-g} \left[ -\frac12 (\widetilde \rho+\chi)^2 g^{\mu\nu}\partial_\mu\theta\partial_\nu\theta -\frac12 g^{\mu\nu}\partial_\mu\chi\partial_\nu\chi -V_{\rm sr}(\theta) -V(\chi) \right] ~,
		\label{eq:qsfi_action}
	\end{equation}
	where the field $\theta$ plays the role of inflaton, and $\chi$ is the isocurvature mode. The potential $V(\chi)$ stabilizes the radial direction and the radius of trajectory is $\rho \equiv \widetilde \rho + \chi_0$, where $\chi_0$ is the effective minimum of the radial field.
	
	The primordial fluctuations along and orthogonal to the trajectory are denoted by $\delta \phi \equiv \rho \delta \theta$ and	$\delta \chi \equiv \chi-\chi_0$, respectively. We expand the radial
	potential in terms of $\delta \chi$,
	\begin{equation}
		V(\chi_0+\delta\chi) = V(\chi_0) + V'\delta\chi + \frac12 V''\delta\chi^2 + \frac16 V'''\delta\chi^3 + \frac1{24}V''''\delta\chi^4 + \cdots ~,
		\label{eq:Vexpand}
	\end{equation}
	where all derivatives are evaluated at $\chi_0$. The perturbed Lagrangian up to quartic order is then ~\cite{Chen:2018xck}
	\begin{equation}
		\begin{split}
			\mathcal{L} ={}& \frac{a^2}{2} \left[ (\delta\phi')^2 - c_\phi^2(\partial_i\delta\phi)^2 + (\delta\chi')^2 - c_\chi^2(\partial_i\delta\chi)^2 \right] - \frac{a^4}{2} m^2 \delta\chi^2 \\
			& +	a^3 \lambda_2 \, \delta\chi \, \delta\phi' - a^4 \left( \frac{\lambda_3}{6}\delta\chi^3 + \frac{\lambda_4}{24}\delta\chi^4 \right) ~.
		\end{split}
		\label{eq:qsfi_general_sound_lagrangian}
	\end{equation}
	Here $c_\phi$ and $c_\chi$ are the propagation speeds of the adiabatic and massive modes and we define $r_s \equiv c_\phi / c_\chi$ for later convenience. We also have
	\begin{equation}
		m^2 = V'' - \dot\theta_0^2 ~,~ \lambda_2 = 2\dot\theta_0 ~,~ \lambda_3 = V''' ~,~ \lambda_4 = V'''' ~.
		\label{eq:qsfi_couplings}
	\end{equation}
	
	In canonical inflation scenarios, the non-Gaussianities are suppressed by the slow-roll condition. Here, the value of couplings $\lambda_3$ and $\lambda_4$ that generate the non-Gaussianity of the massive field are not constrained by slow-roll conditions and can naturally be large. 
	
	The free mode functions of the massive field are
	\begin{equation}
		\delta_k (\tau) \propto (-\tau)^{3/2} H_{\nu}^{(1)} (-c_{\chi} k\tau) ~,~ \nu \equiv \sqrt{9/4-m^2/H^2} ~,
		\label{eq:massive_mode_function}
	\end{equation}
	where $\tau$ is the conformal time and $H_{\nu}^{(1)}$ the first Hankel function. When the isocurvature mode is heavy enough, $m>3H/2$, $\nu$ becomes imaginary and the massive field acts as an underdamped oscillator. For convenience, we define $\widetilde{\nu} \equiv -i\nu$ and the late-time evolution of massive field contains a phase factor $\exp[\pm i\widetilde{\nu}\ln(-k\tau)]$. This phase factor becomes logarithmic in momentum ratios, when transferred to the curvature fluctuations $\zeta$, producing the characteristic cosmological collider signal \cite{Chen:2012ge,Arkani-Hamed:2015bza}.
	
	\subsection{Non-Gaussianities}
	\label{subsec:correlators}
	It is convenient to define the mixed propagator \cite{Chen:2017ryl,Chen:2018xck},
	\begin{equation}
		\mathcal{G}_\pm(k;\tau) = \frac{\pi\lambda_2 H}{8k^3 c_\phi c_\chi^2} I_\pm\left(-c_\chi k\tau; r_s\right) ~,
		\label{eq:mixed_propagator}
	\end{equation}
	\begin{align}
		I_\pm\left(z;r_s\right) \equiv {}& e^{-\pi\mathrm{Im}\nu} z^{3/2} \Bigg\{	H_\nu^{(1)}(z) \left[ e^{-i\pi(1-2\nu)/4} \left(C_\nu^{r_s}\right)^* + B_{\nu\mp}^{r_s}(z)
		\right]	\notag\\
		&\hspace{22mm} + H_{\nu^*}^{(2)}(z) \left[ e^{i\pi(1-2\nu)/4} C_\nu^{r_s} + B_{\nu\pm}^{r_s}(z)	\right] \Bigg\} ~.
		\label{eq:mixed_Ipm}
	\end{align}
	\begin{equation}
		B_{\nu\pm}^{r_s}(z) \equiv -i \int_0^z \frac{dz'}{\sqrt{z'}} H_\nu^{(1)}(z') e^{\mp i r_s z'} ~,
		\label{eq:mixed_Bnu}
	\end{equation}
	\begin{equation}
		C_\nu^{r_s} = \frac{2^\nu}{\pi}r_s^{\nu-1/2} \Gamma\left(\frac12-\nu\right) \Gamma(\nu) {}_2F_1\left( \frac14-\frac{\nu}{2}, \frac34-\frac{\nu}{2}; 1-\nu; \frac{1}{r_s^2} \right) + (\nu \to -\nu) ~.
		\label{eq:mixed_Cnu_r}
	\end{equation}
	
	With these notations, the adiabatic three-point function is
	\begin{align}
		& \quad \left\langle \delta\phi(\bm k_1)\delta\phi(\bm k_2)\delta\phi(\bm k_3) \right \rangle = (2\pi)^3 \delta(\bm k_1 + \bm k_2 + \bm k_3) 2\lambda_3\,{\rm Im} \int_{-\infty}^{0}\frac{d\tau}{(-H\tau)^4} \prod_{i=1}^{3}{\cal G}_+(k_i;\tau) \notag\\
		&= \frac{\pi^3\lambda_2^3\lambda_3} {256Hk_a^3k_b^3c_\phi^3c_\chi^3}{\rm Im}\int_0^\infty\frac{dz}{z^4} I_+\left(\frac{k_a}{q}z;r_s\right) I_+\left(\frac{k_b}{q}z;r_s\right) I_+\left(z;r_s\right) ~,
		\label{eq:inflaton_bispectrum}
	\end{align}
	where $q=\max(k_1,k_2,k_3)$ and the remaining two
	momentum magnitudes are denoted by $k_a$ and $k_b$. Defining 
	\begin{equation}
		\mathcal{L}_n(\{k_i\}) \equiv \left[\prod_{i=1}^{n} \mathcal{P}_\zeta(k_i) \right]^{1/2} ~,
	\end{equation}
	the bispectrum becomes 
	\begin{equation}
		B_\zeta(k_1,k_2,k_3) = (2\pi)^4 \frac{\mathcal{L}_3(k_1,k_2,k_3)}{(k_1k_2k_3)^2}	\left(\frac{\lambda_2}{H}\right)^3 \left(\frac{\lambda_3}{H}\right) \widehat{\mathcal{S}}(k_1,k_2,k_3) ~,
	\end{equation}
	\begin{align}
		\widehat{\cal S}(k_1,k_2,k_3) ={}& -\frac{\pi^2}{512} r_s^3 c_\phi^{-3/2} \frac{q^2}{k_ak_b} {\rm Im}\int_0^\infty\frac{dz}{z^4} \notag\\
		&\times I_+\left(\frac{k_a}{q}z;r_s\right) I_+\left(\frac{k_b}{q}z;r_s\right) I_+\left(z;r_s\right) ~.
		\label{eq:bispectrum_shape}
	\end{align}
	Similarly, the connected trispectrum becomes 
	\begin{equation}
		T_\zeta(\mathbf{k}_1,\mathbf{k}_2,\mathbf{k}_3,\mathbf{k}_4) = (2\pi)^6 \frac{\mathcal{L}_4(\{k_i\})K^3}{(k_1k_2k_3k_4)^3} \left(\frac{\lambda_2}{H}\right)^4	\widehat{\mathcal{T}}(\mathbf{k}_1,\mathbf{k}_2,\mathbf{k}_3,\mathbf{k}_4) ~,
	\end{equation}
	\begin{equation}
		\widehat{\cal T}(\{\bm k_i\}) =	\frac{\pi^3c_\phi^3}{2^{15}} \left[ \frac{4\lambda_4}{\pi}\,t_c	+ \left(\frac{\lambda_3}{H}\right)^2
		(t_s+t_t+t_u) \right] ~,
		\label{eq:trispectrum_kernel}
	\end{equation}
	Here $K=\sum_{i=1}^4k_i$ and we organize the expression of $t$'s in App. \ref{app:tri}.
	
	To gain more understandings on the result, let us use a lognormal parametrization of curvature power spectrum  \cite{Pi:2020otn,Gow:2020bzo}:
	\begin{equation}
		\mathcal{P}_\zeta(k) =	A_\zeta u_\Delta(x) ~,~ u_\Delta(x) \equiv \frac{1}{\sqrt{2\pi} \Delta} \exp\left[ -\frac{\ln^2(x/(k_*R)}{2\Delta^2}	\right] ~,
	\end{equation}
	where $x=kR$, $k_*$ denotes the peak scale, $\Delta$ represents its logarithmic width, and $A_\zeta = \int d\ln k\,\mathcal{P}_\zeta(k)$ is the integrated power. In our numerical evaluations, we set $\Delta = 0.5$ and $k_*R = \sqrt{2}$. We can then single out the dependence of the amplitude via
	\begin{equation}
		\mathcal{L}_n(\{k_i\}) \equiv \left[\prod_{i=1}^{n} \mathcal{P}_\zeta(k_i) \right]^{1/2} = A_\zeta^{n/2} \left[\prod_{i=1}^{n} u_\Delta(k_iR) \right]^{1/2} ~.
	\end{equation}
	Here, the cumulant $S_3$ and $S_4$ appear to scales $A_\zeta^{-1/2}$ and $A_\zeta^{-1}$, in contradiction to $S_n=\mathcal O(\sigma^0)$ from Eq. \eqref{eq:Sn_def}. This is however not a contradiction: $S_n=\mathcal O(\sigma^0)$ holds once the non-Gaussian amplitude is fixed, but here it's not because we didn't fix the coupling constant $\lambda_2$, $\lambda_3$ and $\lambda_4$. That being said,  the cumulant expansion can be valid only for certain combinations of $\lambda$'s and $\sigma$. In light of this, we define the effective non-Gaussianity parameters as
	\begin{equation}
		f_{\mathrm{NL}}^{\mathrm{eff}} \equiv \left(\frac{\lambda_2}{H}\right)^3 \left(\frac{\lambda_3}{H}\right) A_\zeta^{-1/2} ~,
		\label{eq:fNL_def}
	\end{equation}
	\begin{equation}
		\tau_{\mathrm{NL}}^{\mathrm{CI,eff}} \equiv \left(\frac{\lambda_2}{H}\right)^4 \lambda_4 A_\zeta^{-1} ~,~ \tau_{\mathrm{NL}}^{\mathrm{SE,eff}} \equiv	\left(\frac{\lambda_2}{H}\right)^4 \left(\frac{\lambda_3}{H}\right)^2 A_\zeta^{-1} ~,
		\label{eq:tauNL_def}
	\end{equation}
	which have same structure to conventional ones but not related to any observationally defined ones e.g., $f_{\mathrm{NL}}$ or $\tau_{\mathrm{NL}}$. We can thus factorize the cumulants into 
	\begin{equation}
		S_3 = \widehat{\mathcal C}_3 f_{\mathrm{NL}}^{\mathrm{eff}} ~,~ S_4 = \widehat{\mathcal C}_{4,c} \tau_{\mathrm{NL}} ^{\mathrm{CI,eff}} + \widehat{\mathcal C} _{4,e} \tau_{\mathrm{NL}}^{\mathrm{SE,eff}} ~.
	\end{equation}
	where the aforementioned effective parameters absorb the coupling constant and amplitude. Those parameters must be fixed for a certain QSFI setup before using the cumulant expansions. The parameters $\widehat{\mathcal C}$'s collect the remaining dependences, e.g., momentum-space shapes, $r_s$, $\widetilde\nu$, $\Delta$, and $k_*R$; see App. \ref{app:pro} for more details.

	\section{Impact of Heavy-Field Non-Gaussianity on PBH Formation}
	\label{sec:qsfi_pbh_abundance}
	
	With the setup presented in Sec. \ref{sec:qsfi}, we can numerically compute the peak power-spectrum amplitude required to produce a target PBH abundance. We present the result in this section, and the  complementary result, i.e., the abundance enhancement at fixed power-spectrum peak amplitude in App.~\ref{sec:pbh_abundance}. We consider three representative sound-speed ratios $r_s \equiv c_\phi /c_\chi = 0.3, 1, 3$, corresponding to $(c_\phi, c_\chi) = (0.3, 1)$, $(1, 1)$, and $(1, 1/3)$, alongside four mass parameters $\widetilde{\nu} =\mathrm{Im}\nu= 0.1, 0.3, 0.5, 1$.  Meanwhile, we fix $\lambda_2/H=0.38$ and vary the cubic and quartic self-couplings over $\lambda_3/H, \lambda_4 \in [-1, 1]$. The code used to generate the numerical results and figures presented in this work is publicly available~\cite{Li:2026qsfi_pbh_code}.
	
	\subsection{Peak amplitude of the power spectrum required to produce a fixed PBH abundance}
	\label{sec:required_peak_height}
\begin{figure}[t]
		\centering
		\includegraphics[width=\textwidth]{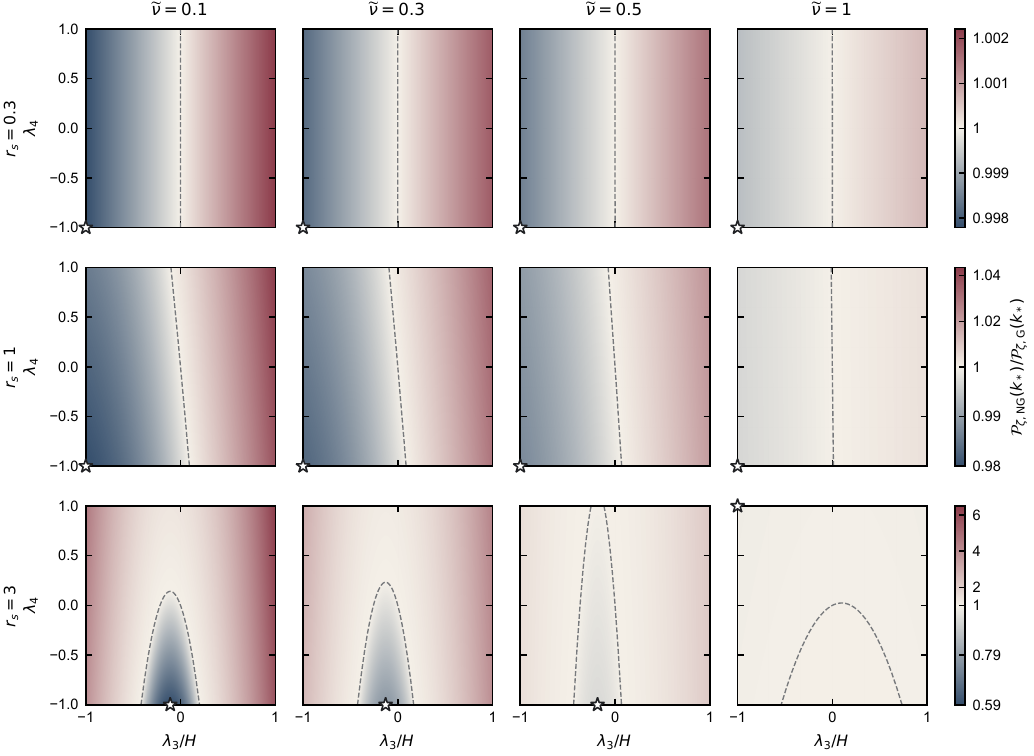}
		\caption{Required power-spectrum peak-amplitude ratio $\mathcal{P}_{\zeta,\mathrm{NG}}(k_*)/\mathcal{P}_{\zeta,\mathrm{G}}(k_*)$ for a target PBH abundance $\beta_\star=10^{-15}$ at $\lambda_2/H=0.38$. Rows correspond to $r_s=0.3$, $1$, and $3$, and columns to $\widetilde{\nu}=0.1$, $0.3$, $0.5$, and $1$. Stars indicate the minimum in each panel, while dashed contours mark $\mathcal{P}_{\zeta,\mathrm{NG}}(k_*)/\mathcal{P}_{\zeta,\mathrm{G}}(k_*)=1$. The perturbative bound $\mathcal Q_{\rm NG} \lesssim6$ is satisfied throughout the displayed parameter space.}
		\label{fig:full-sound-speed-peak-height}
	\end{figure}

	For a target abundance $\beta_* = 10^{-15}$, we solve Eq.~\eqref{eq:beta_NG_S3S4} for the required peak amplitude $\mathcal{P}_{\zeta,\mathrm{NG}}(k_*)$ of the power spectrum and compare it with the Gaussian one $\mathcal{P}_{\zeta,\mathrm{G}}(k_*)$ from Eq.~\eqref{eq:beta_Gaussian}. Following Sec.~\ref{subsec:correlators}, the effective non-Gaussian parameters $f_{\mathrm{NL}}^{\mathrm{eff}}, \tau_{\mathrm{NL}}^{\mathrm{CI,eff}}, \tau_{\mathrm{NL}}^{\mathrm{SE,eff}}$ are kept fixed during this evaluation. Fig.~\ref{fig:full-sound-speed-peak-height} displays the resulting ratio $\mathcal{P}_{\zeta,\mathrm{NG}}(k_*)/\mathcal{P}_{\zeta,\mathrm{G}}(k_*)$, where the coupling values plotted on the axes are defined at $\mathcal{P}_{\zeta,0}(k_*) = 10^{-2}$ using Eqs.~\eqref{eq:fNL_def} and~\eqref{eq:tauNL_def}, with the corresponding minima summarized in Table~\ref{tab:fixed_abundance_extrema}. 
	
	\begin{table}[htbp]
		\centering
		\small
		\begin{tabular}{cccccc}
			\toprule
			$r_s$ & $\widetilde{\nu}$ & $\min\,\mathcal P_{\zeta,\mathrm{NG}}/\mathcal P_{\zeta,\mathrm G}$
			& $\lambda_3/H$ & $\lambda_4$ & $\mathcal Q_{\rm NG}$ \\
			\midrule
			$0.3$ & $0.1$ & $0.997783$ & $-1.000$ & $-1.000$ & $0.0135242$ \\
			$0.3$ & $0.3$ & $0.998183$ & $-1.000$ & $-1.000$ & $0.0110866$ \\
			$0.3$ & $0.5$ & $0.998477$ & $-1.000$ & $-1.000$ & $0.00928744$ \\
			$0.3$ & $1.0$ & $0.999316$ & $-1.000$ & $-1.000$ & $0.00417012$ \\
			\midrule
			$1.0$ & $0.1$ & $0.980063$ & $-1.000$ & $-1.000$ & $0.121596$ \\
			$1.0$ & $0.3$ & $0.983143$ & $-1.000$ & $-1.000$ & $0.102814$ \\
			$1.0$ & $0.5$ & $0.987339$ & $-1.000$ & $-1.000$ & $0.0772234$ \\
			$1.0$ & $1.0$ & $0.996881$ & $-1.000$ & $-1.000$ & $0.0190264$ \\
			\midrule
			$3.0$ & $0.1$ & $0.588146$ & $-0.109$ & $-1.000$ & $2.50238$ \\
			$3.0$ & $0.3$ & $0.776814$ & $-0.129$ & $-1.000$ & $1.35876$ \\
			$3.0$ & $0.5$ & $0.944460$ & $-0.185$ & $-1.000$ & $0.338644$ \\
			$3.0$ & $1.0$ & $0.989042$ & $-1.000$ & $+1.000$ & $0.0668370$ \\
			\bottomrule
		\end{tabular}
		\caption{Minimum required peak-amplitude ratios, their locations in coupling space, and the corresponding $\mathcal Q_{\rm NG}$ values for $\beta_*=10^{-15}$ and $\lambda_2/H=0.38$. All listed points satisfy $\mathcal Q_{\rm NG}\lesssim 6$.}
		\label{tab:fixed_abundance_extrema}
	\end{table}
	
	Tab. \ref{tab:fixed_abundance_extrema} shows a clear dependence on both the sound-speed ratio $r_s$ and the heavy-field mass parameter $\widetilde{\nu}$. This pronounced sensitivity to $r_s$ can be traced to the explicit $r_s^3$ prefactor in the reduced bispectrum in Eq.~\eqref{eq:bispectrum_shape}, as well as the $r_s^5$ prefactors in the contact and scalar-exchange kernels in Eqs.~\eqref{eq:contact_kernel} and~\eqref{eq:exchange_s_kernel}. In addition, $r_s$ affects the mixed propagators through the functions $I_\pm$ in Eq.~\eqref{eq:mixed_Ipm}. As a result, the magnitudes of the projected cumulants increase with $r_s$ and are substantially larger at $r_s=3$ than at $r_s=0.3$ or $1$. By contrast, increasing $\widetilde{\nu}$ ($=\mathrm{Im}\,\nu$) suppresses the heavy-field contribution through the Boltzmann-like factor $e^{-\pi\mathrm{Im}\nu}$ in Eqs.~\eqref{eq:mixed_Ipm} and~\eqref{eq:exchange_s_kernel}. Consequently, the enhancement grows as the heavy-field mass approaches $3H/2$ and as the sound-speed ratio increases.
	
    As a concrete illustration, taking $r_s=3$ and $\widetilde{\nu}=0.1$ with $\lambda_3/H=-0.109$ and $\lambda_4=-1$, we find $\mathcal{P}_{\zeta,{\rm NG}}(k_*)/\mathcal{P}_{\zeta,{\rm G}}(k_*)=0.588146$, namely the required amplitude of power spectrum reduces to roughly half of its original value. This parameter set also gives $\mathcal{Q}_{\rm NG}=2.50238$, safely satisfing $\mathcal{Q}_{\rm NG}\lesssim 6$ and a detailed perturbative analysis is organized in Sec. \ref{sec:perturbative_analysis}.

    The effect of couplings $(\lambda_3/H,\lambda_4)$, on the other hand, are more subtle. For simplicity, we focus on the $(r_s,\widetilde{\nu})=(3,0.1)$ case, where all three projection coefficients are negative, as shown in App.~\ref{app:pro}. According to Eqs.~\eqref{eq:fNL_def} and~\eqref{eq:tauNL_def}, negative cubic and quartic couplings therefore produce positive skewness and a positive contact contribution to the kurtosis, respectively. The contact term thus lowers the required peak-amplitude ratio monotonically as $\lambda_4$ decreases. As a result, the minimum of the power spectrum ratio stays at the scan boundary $\lambda_4=-1$. By contrast, the negative scalar-exchange contribution scales as $(\lambda_3/H)^2$, and its quadratic suppression becomes comparable to the linear skewness enhancement for $\lambda_3/H<0$ owing to the large coefficient at $r_s=3$. This competition of bispectrum and trispectrum contributions leaves a difficulty in deciding the location of the minimum. 
	
	We organize the remaining parameter combinations as follows:
	\begin{itemize}
		\item For $r_s=3$ and $\widetilde{\nu}=0.3$ or $0.5$, the same competition determines the locations of the minimum ratio. All three projection coefficients remain negative, but their magnitudes decrease as $\widetilde{\nu}$ increases. The minima occur at $(\lambda_3/H,\lambda_4)=(-0.129,-1)$ and $(-0.185,-1)$, respectively, and the resulting non-Gaussian correction becomes weaker as $\widetilde{\nu}$ increases.
		
		\item For $(r_s,\widetilde{\nu})=(3,1)$, both trispectrum projection coefficients are positive. A positive quartic coupling and the scalar-exchange contribution therefore both lower the required peak-amplitude ratio, whereas the bispectrum contribution favors $\lambda_3/H<0$. The minimum consequently occurs at $(\lambda_3/H,\lambda_4)=(-1,1)$.
		
		\item For $r_s=0.3$ and $1$, all three projection coefficients are negative for every mass considered. The negative scalar-exchange contribution is negligible because $|\hat{\mathcal{C}}_{4,e}|$ is much smaller than $|\hat{\mathcal{C}}_3|$. The required peak-amplitude ratio decreases monotonically as the couplings become more negative. For every mass considered, the minimum therefore occurs at $(\lambda_3/H,\lambda_4)=(-1,-1)$. 	
	\end{itemize}

	\subsection{Comparison of bispectrum and trispectrum contributions}
	\label{sec:bispectrum_trispectrum_comparison}
	In the above, we find the contributions of bispectrum and trispectrum can be comparable, so it is instructive to examine their individual contributions. From Eq.~\eqref{eq:beta_NG_S3S4}, the relative weighting of the bispectrum and trispectrum in shaping the non-Gaussian tail is determined by $S_3/6$ and $\delta_c S_4/24$. We therefore evaluate the signed value of each contribution and the ratio of their absolute magnitudes: 
	\begin{equation}
		\mathcal{R}_{43}\equiv \left| \frac{\delta_c S_4/24}{S_3/6} \right| = \left| \frac{S_4/8}{S_3} \right|.
		\label{eq:R43_definition}
	\end{equation}
	\begin{figure}[t]
		\centering
		\includegraphics[width=0.55\textwidth]
		{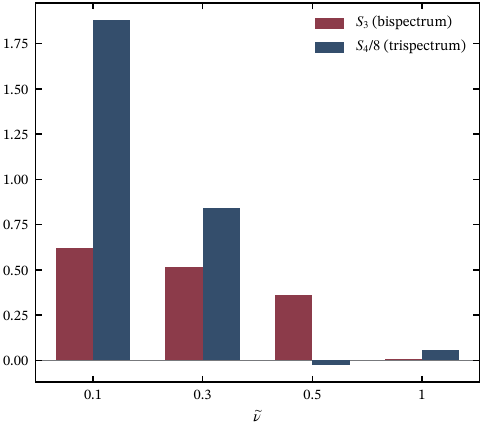}
		\caption{Signed bispectrum and trispectrum contributions at the parameter point minimizing the required peak-amplitude ratio for $r_s=3$, evaluated at $\lambda_2/H=0.38$. The bars show $S_3$ and $S_4/8$ for each mass parameter.}
		\label{fig:r3-relative-contributions}
	\end{figure}
	\begin{table}[t]
		\centering
		\begin{tabular}{cccc}
			\toprule
			$\widetilde\nu$ & $S_3$ & $S_4/8$ & $\mathcal{R}_{43}$ \\
			\midrule
			$0.1$ & $0.622528$ & $1.87985$ & $3.01971$ \\
			$0.3$ & $0.514701$ & $0.844059$ & $1.63990$ \\
			$0.5$ & $0.363592$ & $-0.0249477$ & $0.0686146$ \\
			$1.0$ & $0.00822255$ & $0.0586144$ & $7.12850$ \\
			\bottomrule
		\end{tabular}
		\caption{Signed bispectrum and trispectrum contributions at the fixed-abundance optimal points $r_s=3$, $\lambda_2/H=0.38$, $\beta_\star=10^{-15}$. }
		\label{tab:r3-relative-contribution-summary}
	\end{table}
    For illustrative purpose, we calculate it at the minimum of the peak-amplitude ratio. We also focus on $r_s=3$, where the non-Gaussian correction and the competition between the bispectrum and trispectrum are most pronounced, and show the result in Fig.~\ref{fig:r3-relative-contributions} and Table~\ref{tab:r3-relative-contribution-summary}. For $\widetilde{\nu}=0.1$, the negative quartic coupling produces a positive contact contribution that outweighs the negative scalar-exchange term. The trispectrum contribution is therefore positive, reinforces the skewness contribution, and provides the leading correction to the high-density tail.
	The remaining masses exhibit the following behavior:
	\begin{itemize}
		\item For $\widetilde{\nu}=0.3$, the positive contact contribution also outweighs the negative scalar-exchange term. The trispectrum contribution remains positive and reinforces the skewness contribution.
		
		\item For $\widetilde{\nu}=0.5$, the negative scalar-exchange term becomes dominant, making the trispectrum contribution negative. It therefore partially cancels the bispectrum contribution and skewness becomes the main source of the PBH enhancement.
		
		\item For $\widetilde{\nu}=1$, both trispectrum projection coefficients turn positive, so that the contact and scalar-exchange terms constructively reinforce the bispectrum. The trispectrum dominates the high-density tail correction in this case, although the overall non-Gaussian correction remains small due to the Boltzmann suppression of the heavy field.
	\end{itemize}
	These results demonstrate that including the trispectrum is essential for characterizing the high-density tail in QSFI, as its contribution can be comparable to that of the bispectrum and may even dominate the total non-Gaussian correction.
	
	\subsection{Perturbative Analysis}
	\label{sec:perturbative_analysis}
	Now we perform a detailed perturbative analysis to confirm the robustness of our results. We shall examine three independent directions: 
    \begin{itemize}
        \item The deformation of the constant-turn background, , i.e., whether the perturbation to heavy field $\chi$ is large enough to make it deviates from the constant-turn background assumed in QSFI.
        \item The back-reaction from heavy-field self-interaction, and its back-reaction to the background evolution.
        \item The quadratic transfer mixing, i.e., whether the leading-order transfer approximation between the fluctuations $\delta \chi$ and $\delta \phi$ is valid.
    \end{itemize}
     Since it's not realistic to conduct the perturbative check for all parameter regions, we focus on the point, $r_s=3$, $\widetilde\nu=0.1$, $(\lambda_2/H,\lambda_3/H,\lambda_4)=(0.38,-0.109,-1)$, which makes $\mathcal P_{\zeta,{\rm NG}}(k_*)/\mathcal P_{\zeta,{\rm G}}(k_*)=0.588146$. The stability of this point would be sufficient to support our claim that the heavy field in QSFI scenario can assist PBH formation.
	
	\paragraph{Background deformation.}
	Eq.~\eqref{eq:qsfi_couplings} gives $\dot\theta_0/H=0.19$, $m^2/H^2=2.26$, and $V''/H^2=2.2961$, where $V''$ is evaluated at $\chi_0$. The $\Delta\chi$ denotes the shift of $\chi_0$ relative to the minimum of the radial potential in the absence of turning. For a constant-turn trajectory of radius $\rho$, the radial equation of motion derived from Eq.~\eqref{eq:qsfi_action} imposes $V'(\chi_0)=\rho\dot\theta_0^2$. Expanding $V'$ to linear order about the unturned minimum, where $V'=0$, gives $V'(\chi_0)\simeq V''\Delta\chi$, so that
	\begin{equation}
		\frac{\Delta\chi}{\rho}\simeq\frac{\dot\theta_0^2}{V''}=1.57\times10^{-2} ~.
		\label{eq:background_displacement}
	\end{equation}
	Introducing the slow-roll parameter $\epsilon\equiv-\dot H/H^2=\rho^2\dot\theta_0^2/(2M_{\rm Pl}^2H^2)$, the turning kinetic energy contributes a fraction $\epsilon/3$ of the total density $3M_{\rm Pl}^2H^2$, while the excess radial potential energy $\Delta V\simeq V''(\Delta\chi)^2/2$ contributes
	\begin{equation}
		\frac{\Delta V}{3M_{\rm Pl}^2H^2}
		\simeq \frac{\epsilon}{3}\frac{\dot\theta_0^2}{V''}
		=5.24\times10^{-3}\,\epsilon ~.
		\label{eq:background_energy_backreaction}
	\end{equation}
	The small displacement in Eq.~\eqref{eq:background_displacement} and the slow-roll suppression of both energy fractions in Eq.~\eqref{eq:background_energy_backreaction} show that the turn does not significantly deform the classical background.
	
	\paragraph{Heavy-field self-interactions.}
	At horizon crossing, the root-mean-square (rms) amplitude of the heavy-field fluctuation, $\delta\chi_{\rm rms}\equiv\langle\delta\chi^2\rangle^{1/2}$, is bounded by the conservative massless estimate $\delta\chi_{\rm rms}\lesssim H/(2\pi c_\chi^{3/2})=0.83H$ for $c_\chi=1/3$. The nonzero mass further suppresses the variance. Comparing the nonlinear radial forces from the cubic and quartic terms with the linear restoring force $V''\delta\chi$ at $\delta\chi=\delta\chi_{\rm rms}$ gives
	\begin{equation}
		\frac{|\lambda_3|\delta\chi_{\rm rms}}{2V''}\lesssim1.96\times10^{-2} ~,
		\qquad
		\frac{|\lambda_4|\delta\chi_{\rm rms}^2}{6V''}\lesssim4.96\times10^{-2} ~,
		\label{eq:self_interaction_control}
	\end{equation}
	so both corrections stay at the percent level. The radial potential truncated at quartic order is
	\begin{equation}
		V_{\rm local}(\delta\chi) = \frac12 V''\delta\chi^2 + \frac{\lambda_3}{6}\delta\chi^3 + \frac{\lambda_4}{24}\delta\chi^4~.
		\label{eq:local_potential}
	\end{equation}
	For the parameters here, the nearest stationary point at positive $\delta\chi$ occurs at $\delta\chi\simeq3.55H\simeq4.30\,\delta\chi_{\rm rms}$, supporting the applicability of the local expansion in the regime relevant to our analysis.  To estimate the associated energy backreaction, we denote the energy density stored in the heavy-field fluctuations by $\rho_{\delta\chi}$. Dimensional analysis gives $\rho_{\delta\chi}\sim H^4/(8\pi^2c_\chi^3)$ and hence $\rho_{\delta\chi}/(3M_{\rm Pl}^2H^2)\lesssim\mathcal O(10^{-1})\epsilon$, indicating that the energy backreaction is small. The heavy-field self-interactions are therefore under perturbative control at this point.

	\paragraph{Quadratic transfer mixing.}
	The interaction proportional to $\lambda_2$ in Eq.~\eqref{eq:qsfi_general_sound_lagrangian} transfers the heavy-field fluctuation $\delta\chi$ into the adiabatic fluctuation $\delta\phi$. To test the accuracy of the leading-order transfer approximation adopted in Sec.~\ref{subsec:correlators}, we directly solve the coupled linear mode equations from Eq.~\eqref{eq:qsfi_general_sound_lagrangian} without expanding in $\lambda_2$. Defining $\mathcal P_\phi$ as the resulting power spectrum of $\delta\phi$, $\mathcal P_\phi^{(0)}$ as its value in the decoupled limit, and $\Delta\mathcal P_\phi \equiv \mathcal P_\phi - \mathcal P_\phi^{(0)}$ as the mixing-induced correction, we obtain
	\begin{equation}
		\frac{\mathcal P_\phi}{\mathcal P_\phi^{(0)}}\simeq1.499107 ~,
		\qquad
		\frac{\Delta\mathcal P_\phi}{\mathcal P_\phi^{(0)}}\simeq0.499107 ~.
		\label{eq:quadratic_mixing_correction}
	\end{equation}
	These values remain stable at the percent level under variations of the initial integration time within the subhorizon regime. The bispectrum and trispectrum in Sec.~\ref{subsec:correlators} are calculated at leading order in $\lambda_2$, giving $B_\zeta\propto\lambda_2^3$ and $T_\zeta\propto\lambda_2^4$. The resulting correction cannot be dismissed as negligible, yet it remains subdominant to the leading decoupled contribution and does not signal a breakdown of control in the background or self-interaction sectors. 

    To estimate the impact on the PBH abundance, we make a simple assumption: each $\delta\chi\to\delta\phi$ transfer is enhanced by the same constant factor $\sqrt{Z_\phi}$, where $Z_\phi\equiv\mathcal P_\phi/\mathcal P_\phi^{(0)}$ is the power-spectrum enhancement in Eq.~\eqref{eq:quadratic_mixing_correction}. This gives
	\begin{equation}
		S_3\to Z_\phi^{-1/2}S_3\,,
		\qquad
		S_4\to Z_\phi^{-1}S_4\,.
	\end{equation}
	At the representative point, this lowers $\mathcal Q_{\rm NG}$ from $2.50$ to $1.76$, so the perturbative bound $\mathcal Q_{\rm NG}\lesssim6$ remains satisfied. Repeating the fixed-abundance inversion yields $\mathcal P_{\zeta,\rm NG}(k_*)/\mathcal P_{\zeta,\rm G}(k_*)\simeq0.710$, and at $\mathcal P_{\zeta,0}(k_*)=10^{-2}$ the abundance enhancement becomes $\log_{10}(\beta_{\rm NG}/\beta_{\rm G})\simeq20.6$. The trispectrum-to-bispectrum ratio changes from $\mathcal R_{43}\simeq3.02$ to $2.47$, so the trispectrum remains the leading contribution.
	
	Thus, for the representative point, the background deformation and heavy-field self-interactions, including energy backreaction, are controlled at the percent level. Quadratic transfer mixing is the main source of theoretical uncertainty, but the qualitative conclusions remain unchanged: heavy-field non-Gaussianity still enhances PBH production, and the trispectrum remains important. The leading-order bispectrum and trispectrum predictions used throughout this work are therefore reliable.
	
	\section{Conclusions}
	\label{sec:conclusions}
	In this paper, we have investigated whether primordial non-Gaussianity sourced by heavy fields during inflation, which is widely studied in cosmological collider physics, can also influence primordial black hole formation. To this end, we adopt quasi-single-field inflation (QSFI), a prominent heavy-field model capable of generating both a bispectrum and a sizable trispectrum. By projecting these correlation functions onto the skewness $S_3$ and connected kurtosis $S_4$ of the smoothed density contrast, we connect heavy-field non-Gaussianity to the high-density tail that determines the PBH abundance.
	
	Within the perturbatively controlled regime, heavy-field non-Gaussianity can enhance the high-density tail and thereby increase PBH production. This enhancement becomes stronger as the sound-speed ratio $r_s$ increases and as the heavy-field mass approaches $3H/2$. We further find that the trispectrum contribution can be comparable to the bispectrum contribution or even dominate the total non-Gaussian correction in QSFI scenarios.
	
	Looking ahead, several extensions of this work deserve further attention. First, as discussed in Sec.~\ref{sec:pbh_statistics}, our results rely on peak theory and the Press-Schechter formalism with a fixed threshold $\delta_c$; it would be valuable to reassess the non-Gaussian tail enhancement reported here using the compaction-function approach, which is argued to provide a more robust collapse criterion~\cite{Harada:2015yda,Escriva:2019phb,Yang:2024snb}. Second, while our analysis adopts representative values of $c_\chi$, it would be insightful to investigate what physical mechanisms can realize a smaller heavy-field sound speed and how the resulting PBH abundance is affected. Third, because a target PBH abundance can be reached at a reduced power-spectrum amplitude in the presence of heavy-field non-Gaussianity, it would also be worth quantifying how much this relaxes the CMB $\mu$-distortion bound on $\mathcal{P}_\zeta$~\cite{Fixsen:1996nj,Fixsen:1998kq}. This would widen the viable parameter space for massive PBHs as seeds of supermassive black holes. Furthermore, the correction from quadratic transfer mixing found in Sec.~\ref{sec:perturbative_analysis} motivates a fully resummed calculation of the bispectrum and trispectrum using the exact coupled mode functions. Finally, it would be interesting to examine whether other scenarios going beyond quasi-single-field inflation yield comparable enhancements or distinctive PBH signatures.

\section{Acknowledgements}
We thank Yi-Fu Cai for helpful discussions. MZ is supported by NSFC (Grant No. 12503005), Sichuan Science and Technology Program (Grant No. 2026NSFSC0804), and the Fundamental Research Funds for the Central Universities Grant No. YJ202551. CL is supported by the grant No. UMO-2021/42/E/ST9/00260 from the National Science Centre, Poland.
    
\appendix
	
\section{Expressions of trispectrum and Projection Coefficients}
\label{app:tri}
	
	\subsection{Contact and scalar-exchange kernels}
	The contact kernel takes the form
	\begin{equation}
		t_c = r_s^5\,{\rm Im} \int_0^\infty\frac{dz}{z^4} \prod_{i=1}^4I_+\left(\frac{k_i}{K}z;r_s\right) ~.
		\label{eq:contact_kernel}
	\end{equation}
	The exchange kernels depend on the internal Mandelstam-like momenta
	$k_s=|\bm k_1+\bm k_2|$, $k_t=|\bm k_1+\bm k_3|$, and
	$k_u=|\bm k_1+\bm k_4|$. The $s$-channel contribution is
	\begin{equation}
		\begin{split}
			t_s= &e^{-\pi{\rm Im}\nu}r_s^5 \left(\frac{k_s}{K}\right)^3 {\rm Re} \Bigg[ \int_{\mathcal T\{zz'\}} \left( J_{12s}^{+-}(z)J_{34s}^{++}(z') + J_{34s}^{+-}(z)J_{12s}^{++}(z') \right) \\
			& \hspace{37mm} -\int_{zz'} J_{34s}^{+-}(z)J_{12s}^{-+}(z') \Bigg] ~,
			\label{eq:exchange_s_kernel}
		\end{split}
	\end{equation}
	where
	\begin{align}
		J_{ij\ell}^{\pm+}(z) &=	z^{-5/2} I_\pm\left(\frac{k_i}{k_\ell}z;r_s\right) I_\pm \left(\frac{k_j}{k_\ell}z;r_s\right) H_\nu^{(1)}(z) ~,
		\label{eq:exchange_J_plus} \\
		J_{ij\ell}^{\pm-}(z) &= z^{-5/2} I_\pm\left(\frac{k_i}{k_\ell}z;r_s\right) I_\pm\left(\frac{k_j}{k_\ell}z;r_s\right) H_{\nu^*}^{(2)}(z) ~,
		\label{eq:exchange_J_minus}\\
		\int_{\mathcal T\{zz'\}} &\equiv\int_0^\infty dz\int_0^z dz' ~,~ \int_{zz'}\equiv \int_0^\infty dz\int_0^\infty dz' ~.
		\label{eq:exchange_time_integrals}
	\end{align}
	Here $\ell\in\{s,t,u\}$ labels the exchange channel, and $k_\ell$ denotes the magnitude of the corresponding internal momentum. The $t$- and $u$-channel contributions follow from the substitutions $2\leftrightarrow3$, $k_s\to k_t$ and $2\leftrightarrow4$, $k_s\to k_u$, respectively.
	
	\subsection{Projection coefficients and their numerical values}
	\label{app:pro}
	The bispectrum projection coefficient is defined as
	\begin{align}
		\widehat{\mathcal{C}}_3 \left(r_s,\widetilde{\nu};\Delta,k_*R\right)
		&\equiv
		\frac{2A_\zeta^{2}}{\sigma^4}
		\int_0^\infty \mathrm{d}k_1\,k_1^2
		\int_0^\infty \mathrm{d}k_2\,k_2^2
		\int_{-1}^{1}\mathrm{d}\mu_{12}
		\prod_{i=1}^{3} \left[\mathcal{M}(k_i)W(k_iR)\right]
		\nonumber\\[-2pt]
		&\quad\times
		\frac{\left[\prod_{i=1}^{3}u_\Delta(k_iR)\right]^{1/2}}{(k_1k_2k_3)^2}
		\widehat{\mathcal{S}}(k_1,k_2,k_3).
		\label{eq:C3_projection_coefficient}
	\end{align}
	The contact contribution is determined by
	\begin{align}
		\widehat{\mathcal{C}}_{4,c}\left(r_s,\widetilde{\nu};\Delta,k_*R\right)
		&\equiv
		\frac{\pi c_\phi^3A_\zeta^3}{2^{13}\sigma^6}
		\int_0^\infty\prod_{a=1}^{3}\left(\mathrm{d}k_a\,k_a^2\right)
		\int_{-1}^{1}\mathrm{d}\mu_{12}
		\int_{-1}^{1}\mathrm{d}\mu_{13}
		\int_0^{2\pi}\mathrm{d}\varphi_3
		\nonumber\\[-2pt]
		&\quad\times
		\prod_{i=1}^{4}\left[\mathcal{M}(k_i)W(k_iR)\right]
		\frac{ \left[\prod_{i=1}^{4}u_\Delta(k_iR)\right]^{1/2}K^3}{(k_1k_2k_3k_4)^3}\,t_c ,
		\label{eq:C4_contact_projection_coefficient}
	\end{align}
	and the exchange coefficient is
	\begin{align}
		\widehat{\mathcal{C}}_{4,e}\left(r_s,\widetilde{\nu};\Delta,k_*R\right)
		&\equiv
		\frac{\pi^2c_\phi^3A_\zeta^3}{2^{15}\sigma^6}
		\int_0^\infty\prod_{a=1}^{3}\left(\mathrm{d}k_a\,k_a^2\right)
		\int_{-1}^{1}\mathrm{d}\mu_{12}
		\int_{-1}^{1}\mathrm{d}\mu_{13}
		\int_0^{2\pi}\mathrm{d}\varphi_3
		\nonumber\\[-2pt]
		&\quad\times
		\prod_{i=1}^{4}\left[\mathcal{M}(k_i)W(k_iR)\right]
		\frac{ \left[\prod_{i=1}^{4}u_\Delta(k_iR)\right]^{1/2}K^3}{(k_1k_2k_3k_4)^3}
		\nonumber\\[-2pt]
		&\quad\times
		(t_s+t_t+t_u).
		\label{eq:C4_exchange_projection_coefficient}
	\end{align}
	Here $\mu_{13} \equiv \widehat{\mathbf{k}}_1 \cdot \widehat{\mathbf{k}}_3$, and $\varphi_3$ denotes the azimuthal angle of $\mathbf{k}_3$ around $\mathbf{k}_1$, defined relative to the $(\mathbf{k}_1,\mathbf{k}_2)$ plane.
	The numerical values of these coefficients are listed in Table~\ref{tab:hatted-cumulant-coefficients}. In general, larger sound-speed ratios $r_s$ amplify the magnitudes of all three coefficients, indicating stronger non-Gaussianity. Conversely, increasing the heavy-field mass parameter $\widetilde{\nu}$ monotonically suppresses the magnitudes of all coefficients. In addition, the projection coefficients are strictly negative across nearly all parameter combinations considered here, except at $(r_s, \widetilde{\nu}) = (3, 1)$, where both the trispectrum contact coefficient $\hat{\mathcal{C}}_{4,c}$ and the exchange coefficient $\hat{\mathcal{C}}_{4,e}$ change sign to become positive.
	\begin{table}[t]
		\centering
		\small
		\setlength{\tabcolsep}{4pt}
		\begin{tabular}{ccccc}
			\toprule
			$r_s$ & $\widetilde{\nu}$ & $\widehat{\mathcal C}_3$ & $\widehat{\mathcal C}_{4,c}$ & $\widehat{\mathcal C}_{4,e}$ \\
			\midrule
			0.3 & 0.1 & $-0.027594\pm0.000024$ & $(-3.918\pm0.022)\times10^{-6}$ & $(-7.041\pm0.046)\times10^{-6}$ \\
			0.3 & 0.3 & $-0.022620\pm0.000020$ & $(-3.423\pm0.019)\times10^{-6}$ & $(-5.357\pm0.036)\times10^{-6}$ \\
			0.3 & 0.5 & $-0.018949\pm0.000017$ & $(-2.616\pm0.015)\times10^{-6}$ & $(-3.913\pm0.027)\times10^{-6}$ \\
			0.3 & 1 & $-0.0085081\pm0.0000080$ & $(-8.928\pm0.051)\times10^{-7}$ & $(-1.0507\pm0.0072)\times10^{-6}$ \\
			\midrule
			1 & 0.1 & $-0.36055\pm0.00032$ & $-0.08265\pm0.00047$ & $-0.3477\pm0.0023$ \\
			1 & 0.3 & $-0.29076\pm0.00026$ & $-0.06203\pm0.00035$ & $-0.2529\pm0.0017$ \\
			1 & 0.5 & $-0.20393\pm0.00019$ & $-0.03423\pm0.00019$ & $-0.14352\pm0.00097$ \\
			1 & 1 & $-0.043310\pm0.000043$ & $-0.0012093\pm0.0000072$ & $-0.011796\pm0.000072$ \\
			\midrule
			3 & 0.1 & $-11.652\pm0.010$ & $-10.540\pm0.060$ & $-126.30\pm0.84$ \\
			3 & 0.3 & $-8.1404\pm0.0074$ & $-5.298\pm0.030$ & $-74.47\pm0.49$ \\
			3 & 0.5 & $-4.0098\pm0.0037$ & $-0.7549\pm0.0043$ & $-25.56\pm0.17$ \\
			3 & 1 & $-0.016776\pm0.000028$ & $0.08010\pm0.00046$ & $0.2017\pm0.0015$ \\
			\bottomrule
		\end{tabular}
		\caption{Numerical values of the projected cumulant coefficients for different sound-speed ratios and heavy-field mass parameters.}
		\label{tab:hatted-cumulant-coefficients}
	\end{table}
	
	\section{PBH abundance at fixed power-spectrum peak amplitude}
	\label{sec:pbh_abundance}
\begin{figure}[htbp]
		\centering
		\includegraphics[width=\textwidth]{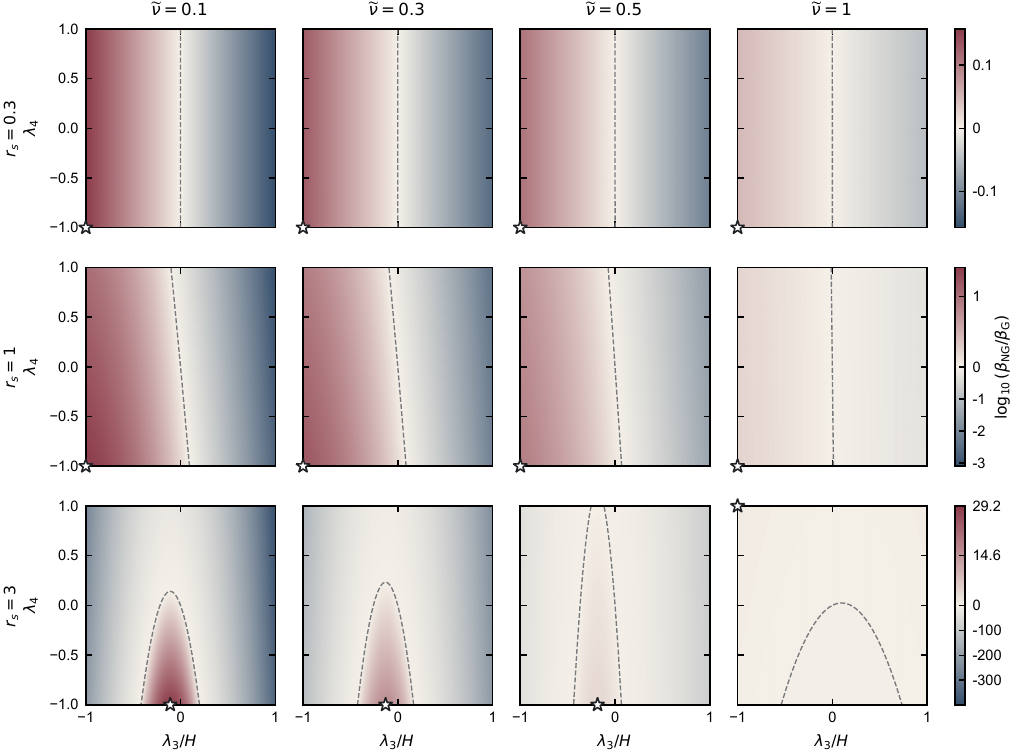}
		\caption{PBH abundance enhancement $\log_{10}(\beta_{\rm NG}/\beta_{\rm G})$ at the reference peak amplitude $\mathcal{P}_{\zeta,0}(k_*)=10^{-2}$ and $\lambda_2/H=0.38$. Rows correspond to $r_s=0.3$, $1$, and $3$, and columns to $\widetilde{\nu}=0.1$, $0.3$, $0.5$, and $1$. Stars indicate the maximum in each panel, while dashed contours mark $\beta_{\rm NG}/\beta_{\rm G}=1$. The perturbative bound $\mathcal Q_{\rm NG} \lesssim 6$ is satisfied throughout the displayed parameter space.}
		\label{fig:full-sound-speed-log-beta}
	\end{figure}

	Fig.~\ref{fig:full-sound-speed-log-beta} shows $\log_{10}(\beta_{\mathrm{NG}}/\beta_{\mathrm{G}})$ for the reference power-spectrum peak amplitude $\mathcal{P}_{\zeta,0}(k_*)=10^{-2}$. The PBH abundance ratio exhibits the same parameter dependence as the required peak-amplitude ratio in Fig.~\ref{fig:full-sound-speed-peak-height}. Specifically, the parameter configurations that minimize the required power spectrum at fixed abundance simultaneously maximize the PBH abundance at fixed power-spectrum peak amplitude.
	Table~\ref{tab:fixed_peak_extrema} summarizes the maximal abundance enhancement and its coupling-space location across the mass and sound-speed parameter space. These results highlight that heavy-field non-Gaussianity can significantly alter the PBH abundance while remaining within the perturbative regime.
	\begin{table}[htbp]
		\centering
		\small
		\begin{tabular}{cccccc}
			\toprule
			$r_s$ & $\widetilde{\nu}$ & $\max\,\log_{10}(\beta_{\rm NG}/\beta_{\rm G})$
			& $\lambda_3/H$ & $\lambda_4$ & $\mathcal Q_{\rm NG}$ \\
			\midrule
			$0.3$ & $0.1$ & $0.157726$ & $-1.000$ & $-1.000$ & $0.0135242$ \\
			$0.3$ & $0.3$ & $0.129297$ & $-1.000$ & $-1.000$ & $0.0110866$ \\
			$0.3$ & $0.5$ & $0.108314$ & $-1.000$ & $-1.000$ & $0.00928744$ \\
			$0.3$ & $1.0$ & $0.0486339$ & $-1.000$ & $-1.000$ & $0.00417012$ \\
			\midrule
			$1.0$ & $0.1$ & $1.41811$ & $-1.000$ & $-1.000$ & $0.121596$ \\
			$1.0$ & $0.3$ & $1.19907$ & $-1.000$ & $-1.000$ & $0.102814$ \\
			$1.0$ & $0.5$ & $0.900615$ & $-1.000$ & $-1.000$ & $0.0772234$ \\
			$1.0$ & $1.0$ & $0.221894$ & $-1.000$ & $-1.000$ & $0.0190264$ \\
			\midrule
			$3.0$ & $0.1$ & $29.1839$ & $-0.109$ & $-1.000$ & $2.50238$ \\
			$3.0$ & $0.3$ & $15.8465$ & $-0.129$ & $-1.000$ & $1.35876$ \\
			$3.0$ & $0.5$ & $3.94942$ & $-0.185$ & $-1.000$ & $0.338644$ \\
			$3.0$ & $1.0$ & $0.779483$ & $-1.000$ & $+1.000$ & $0.0668370$ \\
			\bottomrule
		\end{tabular}
		\caption{Maximum abundance enhancements, their locations in coupling space, and the corresponding $\mathcal Q_{\rm NG}$ values for the reference peak amplitude $\mathcal{P}_{\zeta,0}(k_*)=10^{-2}$ and $\lambda_2/H=0.38$. All listed points satisfy $\mathcal Q_{\rm NG}\lesssim 6$.}
		\label{tab:fixed_peak_extrema}
	\end{table}

	\bibliographystyle{JHEP}
	\bibliography{pbh.bib}

\end{document}